# Adiabatic Invariants in Stellar Dynamics:
# I. Basic concepts

Martin D. Weinberg[1]

Department of Physics and Astronomy

University of Massachusetts/Amherst

## ABSTRACT

The adiabatic criterion, widely used in astronomical dynamics, is based on the harmonic oscillator. It asserts that the change in action under a slowly varying perturbation is exponentially small. Recent mathematical results precisely define the conditions for invariance show that this model does not apply in general. In particular, a slowly varying perturbation may cause significant evolution stellar dynamical systems even if its time scale is longer than any internal orbital time scale. This additional 'heating' may have serious implications for the evolution of star clusters and dwarf galaxies which are subject to long-term environmental forces.

The mathematical developments leading to these results are reviewed, and the conditions for applicability to and further implications for stellar systems are discussed. Companion papers present a computational method for a general time-dependent disturbance and detailed example.

## 1. Introduction

There are two general methods for studying the evolution of stellar systems: 1) self-consistent integration of the equations of motion (e.g. n-body simulation); and 2) solution of the collisionless Boltzmann equation. The first is direct and straightforward in practice. The second solves for the phase-space distribution of orbits rather than the orbits themselves. By Jeans' theorem, the distribution is function of the constants of motion or adiabatic invariants for the system[2]. The evolution is then determined by identifying the those orbits whose invariants are preserved and the change in those that are not.

---

[1] Alfred P. Sloan Foundation Fellow

[2] This assumes that our system may be so characterized, e.g. it is *regular*.

This approach allows the microphysical consequences of $10^5$ to $10^{11}$ orbits to be treated macroscopically and for a galactic age, which is impractical with n-body simulation. This solution of the Boltzmann equation is especially useful when the system has a simple geometry and the perturbation is slower than the orbital times themselves. Since this is true for many scenarios, the adiabatic invariant has become a fundamental tool in stellar dynamics.

The concept of adiabatic invariance is often first encountered as a quantum mechanics student. Indeed, a common example—the one-dimensional pendulum whose length is slowly changing—dates from the 1911 Solvay Congress. This case has been studied extensively, perhaps because it can be, and is the de facto fundamental model for the adiabatic invariant. Although its limitations are well-known (e.g. Bogliubov & Mitropolsky 1961, Kruskal 1962), a better but equally convenient model has not been found. Specifically, a slowly changing perturbation to a harmonic oscillator may be parameterized as a time-varying characteristic frequency for most cases of interest: $\ddot{x} + \omega^2(\epsilon t)x = 0$. This problem may be solved using WKB theory (e.g. Berry and Mount 1972) to show that the change in action is exponentially small in the ratio of the characteristic to perturbation frequency; that is, proportional to $\exp(-\omega/\epsilon)$. This leads to the often used *adiabatic criterion*: if the time scale for change of a perturbation is significantly longer than the characteristic time scale, the action remains invariant. Unfortunately, the perturbed one-dimensional linear oscillator is quite special and does not represent the generic case, as we will see below.

Nonetheless, the standard adiabatic criterion remains widely used because it allows the importance and effect of time dependent perturbations to be easily ascertained. This is especially desirable in the astronomical context where isolated environments rarely exist. In particular, globular clusters suffer strong external perturbations by their embedding galaxies. In addition to the tidal strain and shear, globular clusters are "kicked" when they pass through the disk plane or through the inner galaxy on eccentric orbits. Because the duration of the kick, $\tau$, is small compared to the orbital periods their halo stars, $P$, this is called a *gravitational shock*. Orbits with $P > \tau$ gain energy on average which heats and subsequently expands the cluster. Orbits with $P < \tau$ are assumed to show negligible change by the adiabatic criterion. Accordingly, most researchers confined their attention to impulse approximation (Ostriker et al. 1972). More recently, Chernoff et al.(1986) extended the impulse approximation using Spitzer's (1958) treatment of tidal distortions based on the linear oscillator model.

Using recent results in the theory of classical non-linear systems, this paper will describe why the standard adiabatic criterion based on the oscillator model is false for general systems with more than one degree of freedom and argue that significant changes *can* occur in the adiabatic regime. To begin, we will review the recent progress in adiabatic theorems (§2). Current theory predicts that orbits with periods short compared to the duration of the perturbation, and therefore adiabatically invariant by the standard criterion, may be significantly perturbed nonetheless. Since stellar systems may be viewed as a distribution of non-linear oscillators, the insight from the recent mathematical results help motivate the detailed treatment of gravitational shocking (Paper II) and its application to Fokker-Planck models for globular cluster evolution (Paper III). The implications of the basic mechanism for stellar dynamical systems is discussed in §3.

## 2. Theory of adiabatic invariants

The problem of an integrable Hamiltonian system with a slowly varying perturbation has been studied for the last hundred years, perhaps beginning in earnest with the work of Poincaré. In 1899, Poincaré proved a theorem on the non-existence of integrals of motion in nearly-integrable systems (see Benettin et al.1985a for a discussion and references). The essence of this work is well-known today as the problem of vanishing denominators and the breakdown of canonical perturbation theory. Two relatively recent theorems, the Kolmogorov-Arnold-Moser (KAM) and the Nekhoroshev theorem, partially address these deficiencies of classical canonical perturbation theory. The first shows that most adiabatic invariants are not completely destroyed by the existence of resonances and the second salvages the classical averaging theorem. For systems with many degrees of freedom, we rely on a weaker but more general averaging theorem due to Neistadt (1976). Lochak and Meunier (1988, LM) give a thorough review of the literature on which I will draw heavily.

### 2.1. Adiabatic invariants in one-dimensional systems

#### 2.1.1. Summary of known results

One-dimensional systems are the most exhaustively studied although they are the most restricted in application. Since they are straightforwardly understood analytically and illustrate fundamental concepts, their properties are worth a summary.

Adiabatic theorems may be divided into two classes: periodic and asymptotically autonomous (see Table 1). These classes are not meant to be inclusive but representative of cases that may be precisely defined. A *periodic* perturbation is imposed at some amplitude $\epsilon$ (assumed to be small) for all time. Choosing the unit of time to be the characteristic period, a rough statement of the claim is as follows:

Table 1: One-dimensional adiabatic theorems

|  | Method | |
| Type | Linear | Nonlinear |
| --- | --- | --- |
| Periodic | Lie series (harmonic oscillator) | KAM |
| Asymptotically autonomous | WKB | Nekhoroshev theorem |

> If a perturbed Hamiltonian system which depends on a slowly varying parameter $\tau$, $H = H(p, q, \tau)$ with $\tau \equiv \epsilon t$, has a non-zero bounded frequency, $\Omega = \partial H/\partial I > \Omega_o > 0$ where $\Omega_o$ is a non-zero constant, then the long-term change in action is bounded and of order $\epsilon$.

A stronger version of this claim is proven by the KAM theorem (e.g. Arnold 1978). Although one can think of many mechanical systems in this class, the second class, *asymptotically autonomous* systems, is more relevant astronomically. The claim, which is proven by Nekhoroshev's theorem (see Benettin et al.1984 for an overview and Benettin et al.1985b for details), may be stated as follows:

> Begin with a Hamiltonian depending on a parameter which is slowly varying with time, $H = H(p, q, \lambda(\tau))$, and for which $\lim_{\tau \to \pm\infty} \lambda(\tau)$ exists. If the system can be rewritten in the following form $H = H(p, q, \lambda) = H_0(I, \lambda) + \epsilon H_1(I, \phi, \lambda)$ where $I$ and $\phi$ are the action-angle variables for $H_0$ and whose frequency is bounded as before, $\Omega = \partial H/\partial I > \Omega_o > 0$, then the change in action is bounded and given by $\Delta I \equiv |I(\infty) - I(-\infty)| = \mathcal{O}(e^{-c/\epsilon})$.

The astrophysicist's familiar definition of adiabatic invariant for a perturbation which is slowly "turned on" and "turned off" fits naturally into this class. Note that the invariant is exponentially controlled [that is, its change is $\mathcal{O}(e^{-c/\epsilon})$ with some constant $c$], consistent with one-dimensional harmonic oscillator results of Spitzer (1958). In a one-dimensional system this asymptotic behavior obtains even if the zeroth-order Hamiltonian is non-linear such as for the pendulum. The proof of Nekhoroshev's theorem is based in part on the familiar averaging theorem from canonical perturbation theory (see e.g. Lichtenberg & Liebermann 1983, LL). The overall method of proof will help intuitively motivate the final result and is sketched below.

—– 5 –

### 2.1.2. Sketch of Nekhoroshev's theorem

The averaging principle follows from canonical perturbation theory and is a method to remove the oscillatory dependence on angle variables, from a perturbed Hamiltonian. This is desirable since if the procedure is successful, the new momenta are constants of the motion and the problem is solved.

To summarize this procedure, let us begin with perturbed Hamiltonian of the following form:
$$H = H_0(I,\tau) + \epsilon H_1(I,\phi,\tau) + \cdots. \tag{1}$$
We then attempt to find a canonical transformation to new Hamiltonian, $(I,\phi) \to (\bar{I},\bar{\phi})$, such that the new Hamiltonian independent of $\bar{\phi}$ to first order:

$$\bar{H} = \bar{H}_o + \epsilon \bar{H}_1(\bar{I},\tau) + \cdots. \tag{2}$$

We do this by defining the near identity canonical transformation:
$$S = \bar{I}\phi + \epsilon S_1(\bar{I},\phi,\tau) + \cdots \tag{3}$$

where $S$ is a canonical generating function. To determine $S_1$, substitute the newly generated canonical variables into the Hamiltonian, expand in orders of $\epsilon$ and solve to first order in $\epsilon$. Following LL, this yields:

$$\begin{aligned} \bar{H} &= H_0 + \epsilon \langle H_1 \rangle_{\bar{\phi}} \\ \bar{I} &= I + \epsilon \frac{\{H_1\}_{\bar{\phi}}}{\Omega} \end{aligned} \tag{4}$$

where $\Omega = \partial H_0/\partial \bar{I}$. The notation $\langle H_1 \rangle_{\bar{\phi}}$ denotes the phase-averaged value of $H_1$ at fixed $\tau$ and $\{H_1\}_{\bar{\phi}}$ denotes the rapidly-oscillating phase-dependent part with zero mean. This procedure may be continued to higher order. Most importantly, from equation (4) it is clear that this procedure *only* works for orbital frequencies $\Omega > \Omega_o > 0$ as stated above.

Now because the motion is for a quasiperiodic, the perturbed Hamiltonian may be represented as a Fourier series at fixed $\tau$:

$$\begin{aligned} H_1(I,\phi,\tau) &= \sum_l F_l(I,\tau) e^{il\cdot\phi} \\ &= \sum_{|l|\leq N} F_l(I,\tau) e^{il\cdot\phi} + \sum_{|l|>N} F_l(I,\tau) e^{il\cdot\phi}, \end{aligned} \tag{5}$$

where $l$ is an integer and $N$ will be appropriately chosen below. The analytic part Nekhoroshev's method applies the averaging scheme to the first term in equation (5),

Table 2: Multidimensional adiabatic theorems

| Type | Method |
|---|---|
| Integrable | Neistadt |
| Ergodic | "Thermodynamic" |

yielding an expression of the form

$$\bar{I} = I + \epsilon \sum_{|l| \leq N} \frac{Q_l}{l \cdot \Omega}. \tag{6}$$

Successively applying the averaging theorem leaves only an exponentially-small oscillating remainder for the first term of the Hamiltonian (eq. 5). In addition, one can show that the terms proportional to $\epsilon$ in resulting action series are also exponentially small (Benettin et al.1985b). Finally, assuming that the perturbation is itself analytic, one finds that for sufficiently large $N$, the order of remainder ($|l| > N$) may be estimated using the fact that

$$||F_l|| \sim e^{-\sigma |l|} \tag{7}$$

for some order unity constant $\sigma$; in other words, the Fourier series converges quickly for smooth perturbations. Similar arguments apply for most averaging theorems far from resonance ($|l \cdot \Omega|$ sufficiently far from zero). Much of the full proof is concerned with behavior near resonances, placing limits on the measure of trajectories that linger near a resonance and suffer changes (see LM for details).

### 2.2. Adiabatic invariants in multidimensional systems

#### 2.2.1. Summary of known results

There are many fewer definite results for multidimensional systems; those that exist are for integrable (or nearly integrable) systems or ergodic systems (see Table 2).

Let us first examine the known results, restricting our attention to the integrable case. The main result is an application of Neistadt's averaging theorem:

> Let $\rho$ be a smooth function where $\rho(\epsilon) > 0$ and $c\sqrt{\epsilon} \leq \rho$ for some constant $c$. If the Hamiltonian system $H = H(\mathbf{p}, \mathbf{q}, \lambda(\tau))$ ($\tau = \epsilon t$) is integrable at fixed $\lambda$, then the change in action is bounded and of order $\rho(\epsilon)$ for time periods $t \lesssim 1/\epsilon$ for all but a small measure of initial conditions.

Note that the statement take a similar form to the one-dimensional case above but with the "control parameter", $\rho$ of order greater than $\sqrt{\epsilon}$ rather than $\epsilon$ or $\exp(-c/\epsilon)$.

Why is the adiabatic theorem in multidimensional case so much weaker than in the one-dimensional case? This can be understood by returning to the averaging procedure and discussion including equations (5)–(7) in particular. In the multidimensional case, the method is the same except the actions and angles become vector quantities, $(I, \phi) \Rightarrow (\mathbf{I}, \boldsymbol{\phi})$, the Fourier expansion index $l$ becomes an array of integers $\mathbf{l}$ (cf. eq. 5). As long as $\mathbf{l} \cdot \boldsymbol{\Omega}$ remains non-zero, the method is still valid. However, the simple condition $\Omega_j > \Omega_o > 0$ is no longer sufficient to guarantee that denominators do not vanish: $\mathbf{l} \cdot \boldsymbol{\Omega} = 0$. A commensurability indicates that there is a linear combination of phases which becomes stationary, and a stationary phase is clearly inconsistent with the approximations of the phase-averaging scheme.

Faced with such a commensurability, we may canonically transform our system so that the one of the angle variables is this stationary one and focus separately on this degree of freedom. The perturbed Hamiltonian for one term alone in equation (5) now looks like the non-linear pendulum equation:

$$H = H_0(\bar{I}) + \epsilon A(\tau) \cos \phi. \tag{8}$$

If we expand $H_0$ about the value of $\bar{I}$ at which the corresponding phase becomes stationary, the identification with the pendulum becomes exact: $H = Gp^2/2 - F \cos \phi$ where $G = \partial^2 H_0/\partial \bar{I}^2$, $F = -\epsilon A$ and $p = \bar{I} + c(\tau)$, for some $c$. The value of $p$ is slowly drifting due to the perturbation.

The point of stationary phase in our original Hamiltonian (eq. 5) corresponds to the unstable equilibrium of the pendulum model: the bob standing on its pivot. The unstable trajectory carries the bob from its unstable point around the pivot and back (in an infinite amount of time). If the disturbance is turned off, $A = 0$, the motion is simple rotation with $\dot{\phi}$ =constant. Increasing $A$ from zero, the unstable equilibrium appears but far from this point the pendulum is rotating over its pivot with conserved action, $I = \oint dq \, p/2\pi$, equal to the area under the trajectory in phase plot.

Let us choose a trajectory with non-zero action. As we continue to increase $A$, the unstable trajectory approaches our original trajectory, which until this point has conserved its action. As the unstable trajectory moves through our original trajectory, the pendulum reverses direction of rotation in most cases and changes its action by twice the action of the unstable trajectory. It readily follows from the equations of motion that this change in action is proportional to $\sqrt{F/G}$. Finally returning to our particular term in equation (5), $F = H_1 = \mathcal{O}(\epsilon)$, we have shown that the commensurability causes a change in the action proportional to $\sqrt{\epsilon}$. This is the basic reason for the order of the control parameter $\rho$ in the statement of the multidimensional adiabatic invariant given above. The many problems of astronomical interest, $0.01 \lesssim \epsilon \lesssim 0.3$, so a change of order $\sqrt{\epsilon}$ can not be ignored.

## 3. Discussion

Each orbit in a galaxy or star cluster is a multidimensional nonlinear oscillator. In a spherical system, a star's trajectory is planar with both a radial and azimuthal oscillation. In a triaxial system, an orbit may have three distinct frequencies. A realistic stellar system is an collection of multidimensional nonlinear oscillators whose frequencies are continuously represented in some finite range. The gravitational potential determines the possible range of frequencies and nearly all physically realistic models will have accidental degeneracies where $\mathbf{l} \cdot \mathbf{\Omega} = 0$. Commensurate or nearly commensurate orbits may be strongly affected by a slowly varying external perturbation by the mechanism discussed in §2.2. Formally, a commensurability is only a surface in phase space, but the time-varying amplitude for a realistic perturbation gives the surface finite width. Although many orbits in the system remain invariant, but not all do. Then, averaging over the entire distribution gives a significant contribution, even if $\Omega/\epsilon \gg 1$ everywhere. The overall change to the system can be as strong as for an impulsive perturbation (this will be explicity calculated in the companion papers). The orbits which change communicate this change to the entire system though their contribution to the overall gravitational potential, leading to global evolution.

The conditions for which adiabatic heating is effective are general:

- The model must be nondegenerate. This is not a serious limitation since most systems are nondegenerate. A degenerate system has $\mathbf{l} \cdot \mathbf{\Omega} = 0$ identically for some $\mathbf{l}$ over a large fraction of its phase space. For example, a binary star system is degenerate because $\Omega_1 = \Omega_2$ and $\Omega_3 = 0$. Similarly, the center of a large homogeneous galaxian core has an harmonic potential and therefore $\Omega_1 = \Omega_2 = \Omega_3$.

- The phase space distribution must be smooth and continuous. Again, this is true for most realistic systems. Without this condition, one could conspire to evacuate phase space around the surfaces $\mathbf{l} \cdot \mathbf{\Omega} = 0$ which will eliminate or greatly reduce the adiabatic heating.

Although gravitational shocking of globular clusters by the Galactic disk is emphasized in Papers II and III, there are many other scenarios that may be changed by adiabatic heating. A list describing those currently being investigated follows:

- The globular system appears to contain a distinct thick-disk population as Zinn (1985) has pointed out. The periodic but slow shocking of a thick-disk globular cluster may lead to distinctly different dynamical evolution than the halo population.

- The time-dependent external force felt by a globular on an eccentric orbit is also a gravitational shock, often called a "bulge shock" in its extreme form (e.g. Aguilar et al.1988). However, shock-induced evolution may important for moderate eccentricities due to the enhanced heating in the adiabatic limit leading to increased disruption rates overall.

- Cannibalized dwarf galaxies with high phase space densities are thought to survive tidal disruption in the halo because their stellar orbital frequencies exceed that of their orbit in the 'parent' galaxy. However, the same heating effects may unbind such dwarfs before they reach the core.

- The seminal shocking problem, the response of a star cluster to a passing molecular cloud (Spitzer 1958), may be performed similarly to include the additional heating experienced for slow encounter speeds.

- Similarly, environmental effects of the embedding molecular cloud on protostellar clusters surely cause evolution evolution. This theory allows one to predict simultaneously the initial conditions for those which will survive to be open clusters and the resulting binary star frequencies.

## 4. Summary

Adiabatic invariants are NOT exponentially controlled[3] for *all* orbits if the number degrees of freedom for the system is greater than one. Orbits in a general stellar system have two or three degrees of freedom and therefore some may be strongly perturbed even if the characteristic frequency $\Omega$ is always much larger than the perturbing frequency $\nu$. This leads to measurable heating in the adiabatic regime of satellite galaxies and star clusters. The heating does not depend on any special phase-space conditions other than a continuous distribution function and orbital frequencies which are not everywhere integral multiples of each other; both conditions should obtain for nearly all realistic cases.

---

[3]proportional to $e^{-\Omega/\nu}$ where $\nu$ and $\Omega$ are the perturbation and characteristic orbital frequencies, respectively

Since the adiabatic criterion has been widely used in astronomical dynamics, it is possible that the magnitude of evolution in interacting systems has been underestimated, and significantly so in some cases. Most affected will be heating and disruption rates of relatively small bound subsystems due to both the time-dependent effects of their large-scale orbits and graininess. A general method for applying these ideas computationally will be discussed in the companion paper, and followed up with a detailed example of disk shocking of globular clusters.

I thank David Chernoff, Greg Fahlman, Chigurupati Murali, Doug Richstone and Scott Tremaine for stimulating discussions, and the Institute for Theoretical Physics in Santa Barbara for its hospitality. This work was supported in part by NSF grant PHY89-04035 to ITP and NASA grant NAGW-2224.